\pdfoutput=1

\documentclass[10pt,letterpaper]{article}
\usepackage{graphicx} 
\usepackage[table,xcdraw]{xcolor} 
\definecolor{linkColor}{RGB}{0,0,0} 
\usepackage[T1]{fontenc}
\usepackage[utf8]{inputenc}
\usepackage[english]{babel}
\usepackage{listings}

\usepackage[backend=biber,%
sorting=none,%
style=chem-rsc,%
maxcitenames=2,
maxbibnames=100,
doi=true,
url=true,
articletitle=true,
urldate=long
]{biblatex} 
\renewbibmacro*{addendum+pubstate}{%
  \setunit{\addsemicolon\space}%
  \printfield{addendum}%
  \newunit\newblock
  \printfield{pubstate}}

\usepackage{amsmath,amssymb}
\usepackage{color} 
\usepackage{hyperref}
\hypersetup{hidelinks,%
colorlinks,%
linkcolor=linkColor,%
citecolor=linkColor,%
urlcolor=linkColor}

\usepackage{float}				
\usepackage{booktabs}			
\usepackage{tabu}				
\usepackage[toc,page]{appendix}	
\usepackage{url}				
\usepackage{booktabs}			
\usepackage{wrapfig}			

\usepackage[labelfont=bf,justification=justified]{caption}
\usepackage[labelfont=bf]{subcaption}

\usepackage{multirow}
\usepackage{gensymb}
\usepackage{soul}
\usepackage{sidecap}
\usepackage{siunitx}
\usepackage{placeins}
\usepackage{lipsum}
\usepackage{braket}
\usepackage{upgreek} 
\usepackage{csquotes}

\AtBeginDocument{\RenewCommandCopy\qty\SI}

\usepackage[ruled,vlined,linesnumbered]{algorithm2e}

\let\originalleft\left
\let\originalright\right
\renewcommand{\left}{\mathopen{}\mathclose\bgroup\originalleft}
\renewcommand{\right}{\aftergroup\egroup\originalright}

\usepackage{bm}
\renewcommand{\vec}[1]{\boldsymbol{\mathbf{#1}}} 

\DeclareSIUnit\angstrom{\text {Å}}

\usepackage[a4paper]{geometry}
\usepackage{titlesec}
\usepackage{subfiles} 
\usepackage[shortlabels]{enumitem}

\newcommand{\LSCO}{La\textsubscript{2--$x$}Sr\textsubscript{$x$}CuO\textsubscript{4}}
\newcommand{\BSCCOtwo}{Bi\textsubscript{2}Sr\textsubscript{2}CaCu\textsubscript{2}O\textsubscript{8+$y$}}

\makeatletter
\g@addto@macro\bfseries{\boldmath}
\makeatother

\usepackage[affil-it]{authblk}

\usepackage[en-GB]{datetime2}
\DTMlangsetup[en-GB]{ord=raise}
\newcommand{\dateOfTheReport}{\DTMdisplaydate{2024}{3}{27}{-1}}

\title{Existence and role of low energy charge-paramagnon modes in the strange metal phase of \BSCCOtwo}
\author[1,*]{C. F. J. Flipse}
\author[1]{T. J. N. van Stralen}
\author[2]{D. M. Kepaptsoglou}
\author[2]{Q. M. Ramasse}

\affil[1]{Department of Applied Physics, Eindhoven University of Technology, Groene Loper 19, 5612 AP Eindhoven, The Netherlands}
\affil[2]{SuperSTEM Laboratory, SciTech Daresbury Campus, Daresbury WA4 4AD, United Kingdom}
\affil[*]{Corresponding author: \href{mailto:C.F.J.Flipse@tue.nl}{C.F.J.Flipse@tue.nl}}

\date{\dateOfTheReport}

\begin{document}

\maketitle
\thispagestyle{empty}

\section*{Abstract}
The strange metal phase is characteristic for the $T$-linear dc resistivity behaviour over a large $T$-range. The effect of the strength of the charge-paramagnon interactions on the charge fluctuations in optimally doped and underdoped regions of \BSCCOtwo\ (Bi-2212) may shine light on the anomalous behaviour of the optical conductivity response in the energy range $50 - 500 \, \textrm{meV}$. We present a preliminary analysis of a single initial run of electron energy loss spectroscopy (EELS) measurements done in a scanning transmission electron microscope (STEM) which exhibit linear dispersive modes separated by $50 \, \textrm{meV}$ energy gaps up to 250 meV in optimally doped Bi-2212. Our observations show similarities with the fluctuating stripes as predicted by Zaanen.

\restoregeometry 
\pagenumbering{arabic}

\section{Introduction} \label{ch:introduction}

\noindent
Can collective charge responses in cuprate superconductors be indicative for the strange metal behavior? How may we understand the mysterious transport phenomena such as the linear-in-temperature resistivity in optimally doped cuprates (“strange metal phase”), reflecting a non-Fermi liquid and characterized by dense many body quantum entanglement? This transport behavior may be described by the laws of hydrodynamics for strongly collective flows and show off as a strongly “incoherent” transport \cite{Marel2003,Marel2006}.\\

\noindent
Hydrodynamics accurately describes most liquids at long length scales compared to the particle-particle mean-free path, but is usually irrelevant for electronic liquids in solids. A hydrodynamic description is based on the existence of variables associated with conserved quantities, while neither the momentum nor the energy of the electron liquid in a solid is conserved. Momentum conservation is violated via various scattering processes, such as electron-impurity, electron-phonon and umklapp scattering. So, even in relatively clean systems such as the electron gas in a semiconductor, the kinetics are typically described by the Boltzmann equation and the conductivity is determined by the corresponding momentum relaxation lengths.\\

\noindent
However, for strongly correlated electron fluids in which the electron-electron mean-free path is small compared to the length scales over which momentum conservation is violated, umklapp scattering is negligible 
and the electron fluid attains local equilibrium on a length scale which is short compared to the scales at which the conservation laws break down, but characterizing hydrodynamic charge transport in strongly coupled systems is a challenging problem. Lately, an explosion in activity in studying large-$N$ systems, such as Sachdev-Ye-Kitaev (SYK) models or AdS2 holography in studying possible connections between transport properties and chaos \cite{Blake2016,Patel2017}. Studying charge-charge interactions can give an insight into this complex matter.\\

\noindent
Upon doping Mott insulators, symmetry breaking leads to the formation of stripes, patterns of organised charge carriers and their spins \cite{Zaanen1989}. State-of-the-art numerical calculations have shown that stripes maintain their characteristic periodicity in a fluctuating form and may exist up to high temperature, as was reported by \citeauthor{Huang2017} \cite{Huang2017}.\\

\noindent 
Measuring the dynamic charge response, revealing the collective charge properties, may elucidate if the imaginary part of this response resembles the predicted anomalous attenuation of plasmons due to the decay into the quantum critical continuum or if for instance other mechanisms are at work. Until now, momentum-resolved electron energy loss spectroscopy (EELS) results by \citeauthor{Mitrano2018} reveal a featureless, non-dispersing plasmon at $1 \, \textrm{eV}$ \cite{Mitrano2018,Husain2019}, which is in contrast to transmission EELS results of \citeauthor{Nucker1989} \cite{Nucker1989,Nucker1991}. In this study, we use a state-of-the-art scanning transimssion electron microscope (STEM) to perform EELS measurements, giving insight into the collective excitations in the low-energy region.

\section{Results} \label{ch:results}

\subsection{Experimental results \& Discussion}

\noindent
All measurements are done on optimally doped \BSCCOtwo\ (Bi-2212), a cuprate superconductor with a superconducting transition temperature $T_\textrm{c}$ of about 91 K. This material contains two pairs of closely-spaced CuO$_2$ planes per unit cell with a Cu--Cu distance of $3.8 \, \si{\angstrom}$, with layers of various compositions in-between. The unit cell of Bi-2212 is typically defined with in-plane lattice parameters $a$ and $b$ which are a factor $\sqrt{2}$ larger than the Cu--Cu distance, i.e. $a = b = 5.4 \, \si{\angstrom}$ due to the presence of an incommensurate ($\sim 4.7 b$) supermodulation in the direction $45\degree$ to the Cu--O--Cu bond. The out-of-plane lattice parameter is $c = 30.8 \, \si{\angstrom}$, though the distance $\ell = c/2$ between two pairs of closely-spaced CuO$_2$ planes is typically referred to. Finally, the reciprocal lattice units are defined as $a_i^* = 2\pi/a_i$, with $a_i$ denoting one of the three lattice parameters.\\

\begin{figure}[H]
    \centering
    \captionsetup{width=0.9\textwidth}
    \includegraphics[width=0.3\textwidth, angle=45]{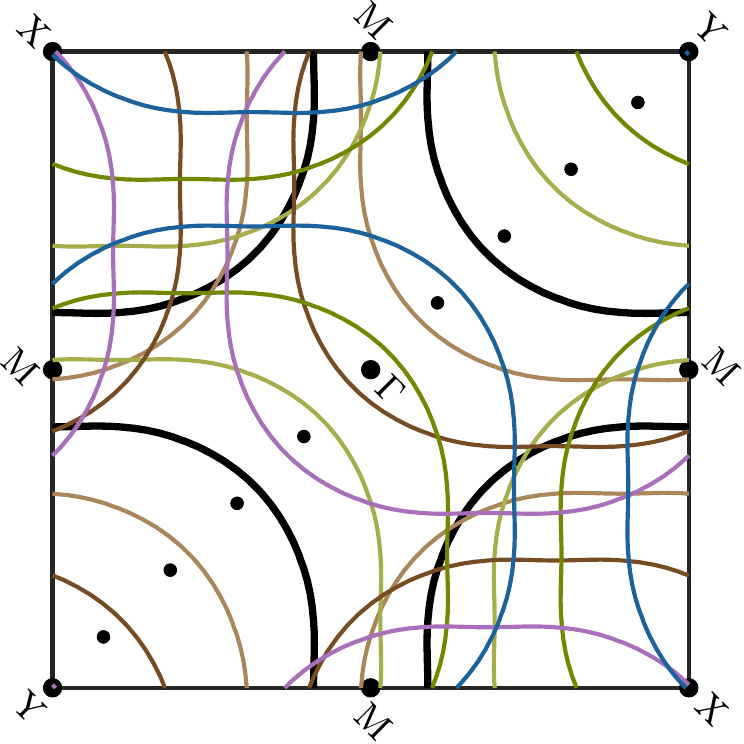}
    \caption{Schematic depiction of the Fermi surface of Bi-2212, including the replicas which originate from the \protect{$4.7b$} supermodulation. The Fermi surface is drawn using the tight binding parameters used by \protect{\citeauthor{Norman1995}} \protect{\cite{Norman1995}}.}
    \label{fig:FermiSurfaceBi2212_final}
\end{figure}

\noindent
A schematic depiction of the Fermi surface of Bi-2212 is given in Fig.~\ref{fig:FermiSurfaceBi2212_final}. It consists of a `main' surface, indicated by the black curve, centered around the $\textrm{X}$ and $\textrm{Y}$ high symmetry points, as well as several Fermi surface replicas, each of which is shown in a different color. These replicas originate from the $4.7b$ supermodulation and each replica is thus centered around a satellite diffraction peak, which are indicated as black dots.\\

\noindent
The electron energy loss spectroscopy (EELS) measurements were performed in a scanning transmission electron microscope (STEM) operated at room temperature. The incoming electron energy is $60 \, \textrm{keV}$ and the energy resolution is $15 \, \textrm{meV}$, as determined by the full width at half maximum (FWHM) of the zero-loss peak. In a STEM, the incoming electron beam is focused onto a small region on the sample with a size of $\Delta x$, which leads to a finite momentum resolution $\Delta q$ by virtue of the Heisenberg uncertainty principle. By tuning the electron optics of the STEM, a balance can be sought between spatial resolution and momentum resolution; see Supplementary \ref{supplementary:EELS_data_acquisition} for more detailed information regarding the STEM setup and the resolutions.\\

\noindent
In this paper, we subdivide the momentum transfer $\vec{q}$ into two components, $q_\parallel$ and $q_\perp$, which denote the momentum transfer along the CuO$_2$ planes ($ab$ plane) and perpendicular to them ($c$ axis), respectively.\\

\begin{figure}[!ht]
\centering
    \captionsetup{width=0.9\textwidth}
\begin{subfigure}{.45\textwidth}
  \centering
  \includegraphics[width=0.95\linewidth]{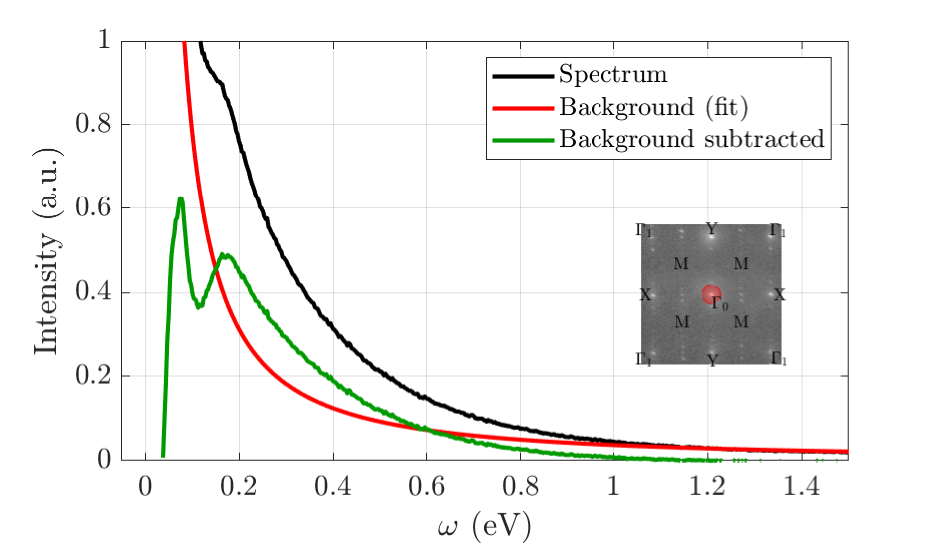}
  \caption{}
  \label{subfig:STEM_2mrad_on_withDiffr}
\end{subfigure}
\begin{subfigure}{.45\textwidth}
  \centering
\includegraphics[width=0.95\linewidth]{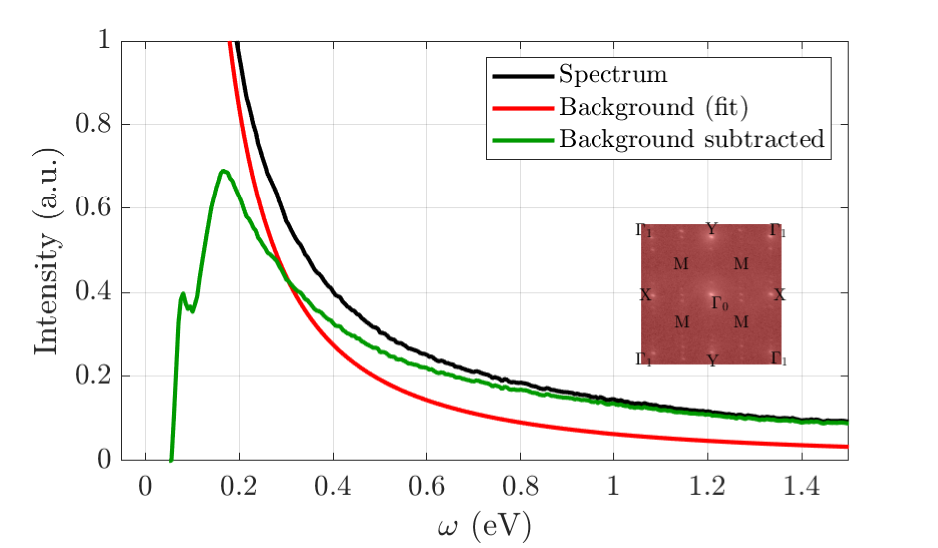}
  \caption{}
  \label{subfig:STEM_30mrad_on_withDiffr}
\end{subfigure}
\caption{Two STEM-EELS measurements on optimally doped Bi-2212. Both measurements are performed along the optical axis (\protect{$q_\parallel = 0$}), but the focusing of the electron beam is different. Inset: Diffraction pattern of Bi-2212, where the red circle depicts the measured area in momentum space and has a diameter \protect{$\Delta q$}. \textbf{(a)} A weakly convergent beam is used, with \protect{$\Delta q \approx 0.37 \, \si{\angstrom}^{-1}$} and a \protect{$1.2 \, \textrm{nm}$} probe size. \textbf{(b)} A highly convergent beam is used, with \protect{$\Delta q \approx 6.9 \, \si{\angstrom}^{-1}$} and a \protect{$0.8 \, \si{\angstrom}$} probe size.}
\label{fig:initial_measurements}
\end{figure}

\noindent
The two spectra in Fig.~\ref{fig:initial_measurements} are obtained in STEM-EELS measurements along the optical axis, hence $q_\parallel = 0$, but with a different focus of the incoming electron beam. The spectrum in Fig.~\ref{subfig:STEM_2mrad_on_withDiffr} is obtained using a weakly convergent beam with a $0.37 \, \si{\angstrom}^{-1}$ momentum resolution and a probe size of $1.2 \, \textrm{nm}$. The spectrum in Fig.~\ref{subfig:STEM_30mrad_on_withDiffr} is obtained using a highly convergent beam with a $6.9 \, \si{\angstrom}^{-1}$ momentum resolution and a probe size of $0.8 \, \si{\angstrom}$. In these graphs, a power law background fit is done to be able to resolve the low energy peaks. Both measured EEL spectra contain the known optical phonons of Bi-2212 at $50$ and $80 \, \textrm{meV}$ \cite{Mills1994}, although the individual phonon peaks in Fig.~\ref{subfig:STEM_2mrad_on_withDiffr} nearly merge and they are not visible in Fig.~\ref{subfig:STEM_30mrad_on_withDiffr} for the chosen background subtraction. Aside from the phonons, a pronounced peak is found in both spectra around $0.2 \, \textrm{meV}$. Furthermore, the well-known $1 \, \textrm{eV}$ plasmon is only present in Fig.~\ref{subfig:STEM_30mrad_on_withDiffr}, although the intensity of this peak is very low. The plasmon mode at $1 \, \textrm{eV}$ was previously reported in several studies \cite{Nucker1989,Nucker1991,Wang1990,Bozovic1990,Levallois2016}, whereas the $0.2 \, \textrm{eV}$ peak has to our knowledge not been observed or reported so far in EELS experiments.\\

\subsubsection{Dispersion low energy modes}

\begin{figure}[H]
    \centering
    \captionsetup{width=0.9\textwidth}
  \includegraphics[width=0.4275\linewidth]{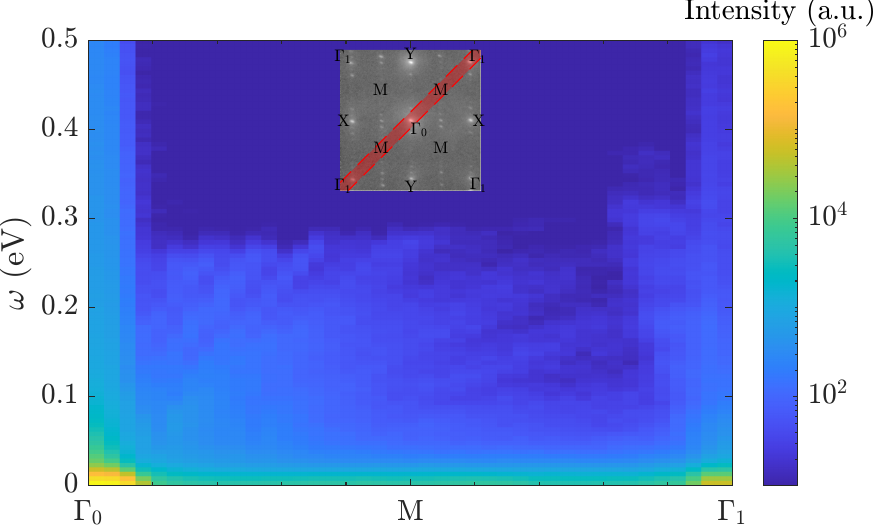}
    \caption{Raw spectra obtained in STEM-EELS measurements for several values of the in-plane momentum \protect{$q_\parallel$} along the anti-nodal direction, where the \protect{$\Gamma_0$}--\protect{$\textrm{M}$} distance is equal to \protect{$a^* / \sqrt{2} \approx 0.83 \, \si{\angstrom}^{-1}$}. Inset: Diffraction pattern of Bi-2212, where the red rectangle illustrates the measurement direction in reciprocal space and has a width $\Delta q$.}
    \label{fig:antinodal_measurements}
\end{figure}

\noindent
The next logical step is to investigate whether the mode at $0.2 \, \textrm{eV}$ displays any dispersion. To this end, we replace the circular aperture with a rectangular slot aperture. In all measurements which are discussed in the remainder of the main text, the electron optics are tuned such that the momentum resolution is roughly $0.23 \, \si{\angstrom}^{-1}$ and the probe size is $1.6 \, \textrm{nm}$. We measured the dispersion in three different directions in momentum space, namely the $\Gamma_0$--$\textrm{Y}$ direction (parallel to the supermodulation), the $\Gamma_0$--$\textrm{X}$ direction (perpendicular to the supermodulation peaks) and the $\Gamma_0$--$\textrm{M}$ direction. In the language of cuprates, these former two correspond to the nodal direction, while the latter refers to the anti-nodal direction. We find that the $0.2 \, \textrm{eV}$ mode only shows a dispersion in the anti-nodal direction. The measured dispersion is illustrated in Fig.~\ref{fig:antinodal_measurements}, where the EEL spectra for $q_\parallel$ values along $\Gamma_0$--$\textrm{M}$--$\Gamma_1$ are stacked side-by-side. Here, several linear-like dispersing branches are visible, each with a slope which corresponds to the group velocity $v_\textrm{g}$. It appears that the modes with high $v_\textrm{g}$ are cut off above $0.25 \, \textrm{eV}$, while the modes with a lower $v_\textrm{g}$ do not reach this cut-off energy and continue all the way up to $\Gamma_1$. Towards $\Gamma_1$, these latter branches bend downwards. Furthermore, the dispersion has apparent $0.05 \, \textrm{eV}$ gaps in-between successive branches and seems disconnected from the branches which start at $\Gamma_0$.\\

\noindent
We want to point out that an expected mirror symmetry around the Brillouin zone boundary at the $\textrm{M}$ point is not observed, which is peculiar since $\Gamma_0$ and $\Gamma_1$ are equivalent points in the Brillouin zone. We will comment on this later in the context of our model in Section \ref{subsec:our_model}.\\

\begin{figure}[H]
    \centering
    \captionsetup{width=0.9\textwidth}
  \includegraphics[width=0.4275\linewidth]{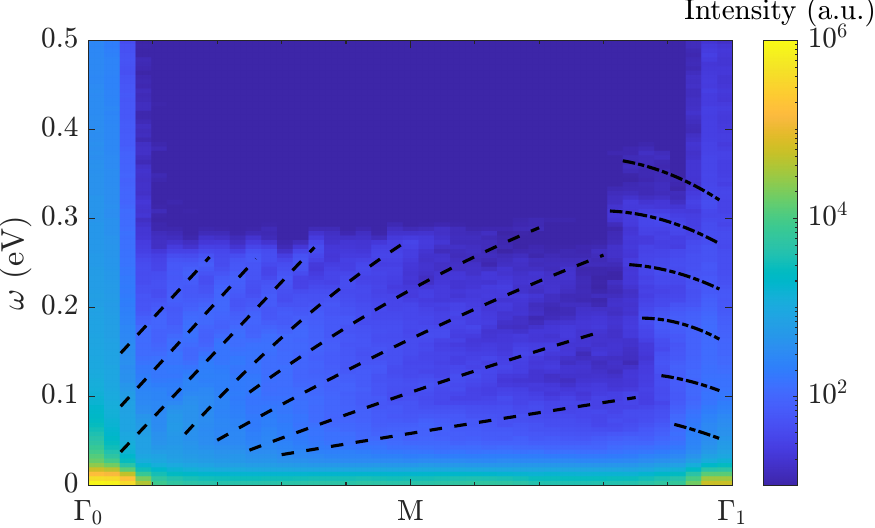}
    \caption{Raw spectra obtained in STEM-EELS measurements for several values of the in-plane momentum \protect{$q_\parallel$} along the anti-nodal direction, including dashed lines which act as a guide to the eyes for the several modes which are found.}
    \label{fig:antinodal_measurements_withLines}
\end{figure}

\noindent
The linear dispersion of the top three measured modes are characteristic of an acoustic plasmon. So far, acoustic plasmons have not been observed in Bi-2212 with EELS, although other techniques like resonant inelastic X-ray spectroscopy (RIXS) have found signatures of acoustic plasmons in similar cuprates \cite{Hepting2018,Nag2020,Singh2022}. Furthermore, the presence of acoustic plasmons in layered materials like cuprates has been predicted via electron gas theory \cite{Fetter1974,DasSarma1982} and the AdS-CFT correspondence (holography) \cite{Mauri2019}. Recently, new holographic calculations on Bi-2212 predict the presence of two acoustic plasmon modes, where the second mode arises due to the presence of two closely-spaced CuO$_2$ planes in Bi-2212 \cite{Eede2023}. All the aforementioned acoustic plasmons only appear if the out-of-plane momentum $q_\perp$ is nonzero. This condition is fulfilled in our STEM-EELS measurements in Fig.~\ref{fig:antinodal_measurements} due to the relatively poor momentum resolution $\Delta q \approx 0.23 \, \si{\angstrom}^{-1}$ which is equivalent to $\sim 1.1 c^*$. The predicted group velocity of these acoustic plasmons is much higher than in our measurements. Namely, the minimum group velocity of the two modes found by \citeauthor{Eede2023} is approximately $5.5$ and $12 \times 10^{5} \, \textrm{m} \, \textrm{s}^{-1}$ when $q_\perp = \pi/\ell$, while the group velocity of the linear modes found in our STEM-EELS experiments is $0.72 \times 10^{5} \, \textrm{m} \, \textrm{s}^{-1}$. Evidently, the group velocity of the measured modes is much smaller than the calculated acoustic plasmon group velocity. So, the plasmons which are calculated by theory are unlikely candidates to explain the dispersion curves in Fig.~\ref{fig:antinodal_measurements_withLines}.\\

\begin{figure}[!ht]
\centering
    \captionsetup{width=0.9\textwidth}
\begin{subfigure}{.45\textwidth}
  \centering
\includegraphics[width=0.95\linewidth]{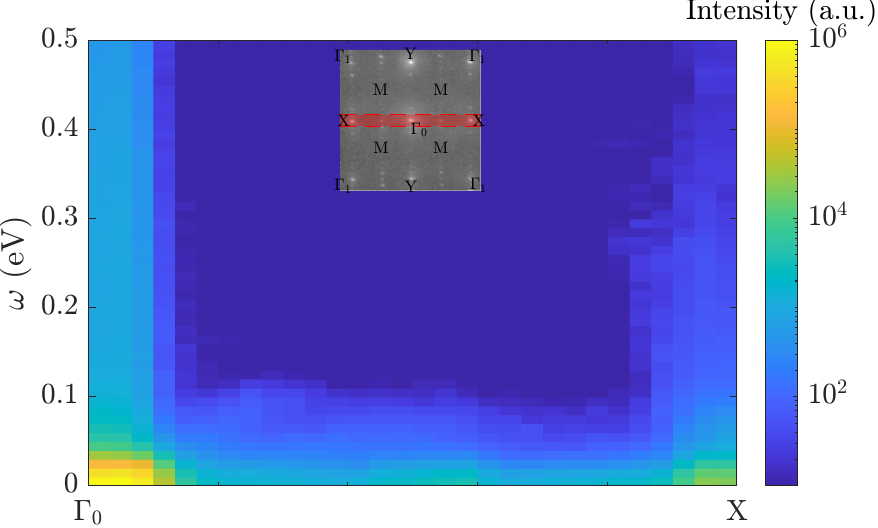}
  \caption{}
  \label{subfig:EELS_100c_withDiffraction}
\end{subfigure}
\begin{subfigure}{.45\textwidth}
  \centering
  \includegraphics[width=0.95\linewidth]{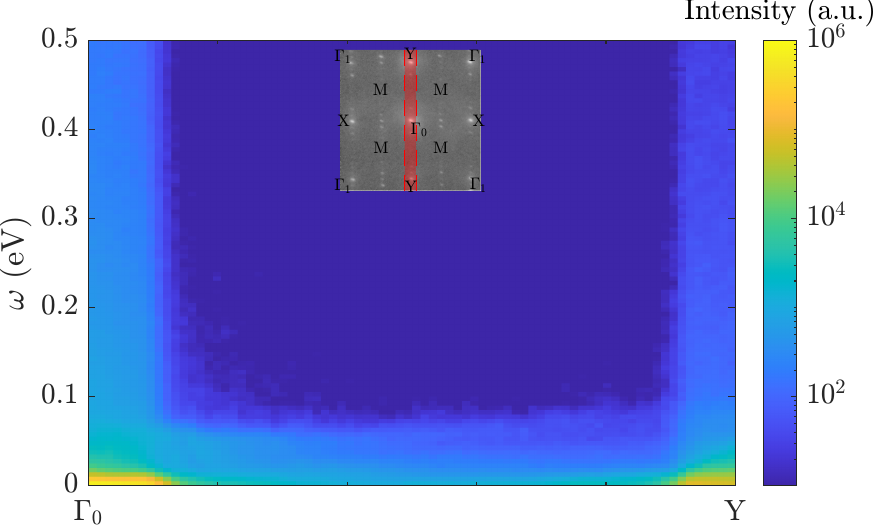}
  \caption{}
  \label{subfig:EELS_010c_withDiffraction}
\end{subfigure}
\caption{Raw spectra obtained in STEM-EELS measurements for several values of the in-plane momentum \protect{$q_\parallel$} along the nodal direction, where the \protect{$\Gamma_0$}--\protect{$\textrm{X}$} and \protect{$\Gamma_0$}--\protect{$\textrm{Y}$} distances are equal to \protect{$a^* \approx 1.17 \, \si{\angstrom}^{-1}$}. The raw spectra are shown for the direction \textbf{(a)} perpendicular (\protect{$\Gamma_0$}--\protect{$\textrm{X}$}) and \textbf{(b)} parallel (\protect{$\Gamma_0$}--\protect{$\textrm{Y}$}) to the \protect{$4.7b$} incommensurate supermodulation. Inset: Diffraction pattern of Bi-2212, where the red rectangle illustrates the measurement direction in reciprocal space and has a width $\Delta q$.}
\label{fig:nodal_measurements}
\end{figure}

\noindent
For the sake of completeness, the dispersion along the nodal direction is shown in Fig.~\ref{fig:nodal_measurements}. These spectra were measured in two different directions in momentum space, namely perpendicular and parallel to the incommensurate supermodulation. Evidently, the modes that are visible in the anti-nodal direction do not appear to be present here, aside from the $50$ and $80 \, \textrm{meV}$ phonons.\\

\noindent
An alternative route for understanding the STEM-EELS spectra of optimally doped Bi-2212 in the anti-nodal direction is the role of (para)magnons which may affect the dynamic charge density response behavior. RIXS experiments on underdoped and optimally doped cuprates have shown the presence of broad, sub-$0.5 \, \textrm{eV}$ magnetic excitations, namely magnons and paramagnon, with a dispersion in the anti-nodal direction \cite{Braicovich2010,Hepting2022,Guarise2014,Dean2013_PRL,Markiewicz2007}. In particular, \citeauthor{Guarise2014} calculated the charge susceptibility for underdoped Bi-2212 by coupling to these (para)magnons via strong magnetic correlations (see Fig.~5c of \cite{Guarise2014}). The existence of these linear modes in their calculations is the only feature which is consistent with our STEM-EELS measurements, as with the holography calculations from \citeauthor{Eede2023}. However, other features of the dispersion curves in Fig.~\ref{fig:antinodal_measurements_withLines} are not obtained in these calculations, such as the cut-off at $0.25 \, \textrm{eV}$ and the correct slope $v_\textrm{g}$. Still, signatures of the dispersion near $\Gamma_1$ might be found in the calculations from \citeauthor{Guarise2014}, in particular in Fig.~5d from \cite{Guarise2014} which represents the contribution of collective spin excitations to the charge.\\

\noindent
The curved dispersive modes at $\Gamma_0$--$\textrm{M}$ at lower energies than the linear ones, may represent charge propagating modes interacting with CDWs of the pseudogap regions \cite{Li2020,Wise2008}.

\subsection{Model} \label{subsec:our_model}

\begin{figure}[H]
    \centering
    \captionsetup{width=0.9\textwidth}
    \includegraphics[width=0.55\textwidth]{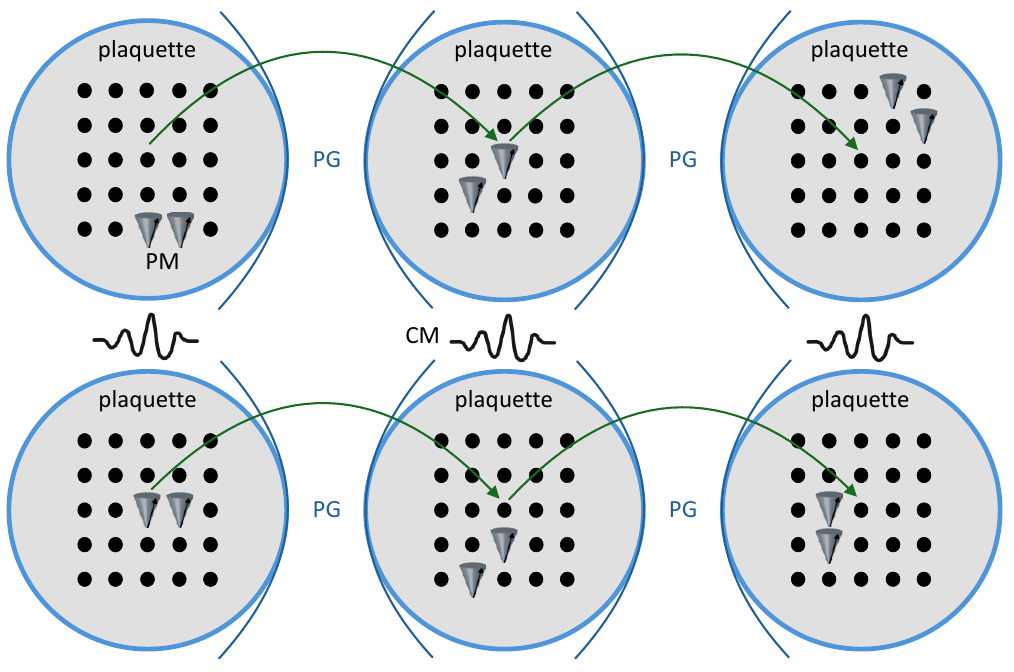}
    \caption{Schematic of the model we propose. The plaquettes are shown schematically as a circle with \protect{$5\times 5$} lattice sites. These plaquettes are in the optimally doped regime, while the region in-between the plaquettes is in the pseudogap (PG) regime. Moreover, paramagnons (PMs) and charge modulations (CMs) are present.}
    \label{fig:CP_model_final_5x5}
\end{figure}

\noindent
We propose the following model, schematically shown in Fig.~\ref{fig:CP_model_final_5x5}, to explain the measured dispersion curves, including the collapse of the charge dispersion at $0.25 \, \textrm{eV}$. Six individual plaquettes are drawn, each containing 25 lattice sites. They form a smectic electronic phase region at $6 \, \textrm{K}$, as shown by \citeauthor{Wise2008} in Fig.~1 of their paper \cite{Wise2008}. These plaquettes represent the optimally doped regions (yellow circular areas in Fig.~1c \cite{Wise2008}), surrounded by regions which are in the pseudogap phase. Paramagnons are present inside these plaquettes and the main issue to address is the unexpected measured charge responses.\\

\noindent
In this model, we expect that the three linear modes near $\Gamma_0$ in Fig.~\ref{fig:antinodal_measurements_withLines} represent the charge fluctuations across the optimally doped plaquettes. Charge fluctuations within a single plaquette should show an almost dispersionless behavior. The existence of paramagnons within the plaquettes can be made plausible by considering evidence of signatures of mutual influence of charge and spin fluctuations from RIXS experiments \cite{Dean2013_PRB,daSilvaNeto2018}. Furthermore, numerical simulations by \citeauthor{Huang2017} show that charge carriers and their spins organize into striped patterns up to high temperatures \cite{Huang2017}. In optimally doped regions, these stripes are relatively small compared to regions in the pseudogap phase \cite{Huang2017}, so the coherence length of paramagnons in the stripes reduces and consequently the number of degrees of freedom of the charge mode increases. Consequently, coherent charge behavior (Drude-like) may persist up to $0.25 \, \textrm{eV}$, which allows for the dispersive modes below $0.25 \, \textrm{eV}$ which we find in the STEM-EELS data.\\

\noindent
In the $\Gamma_1$--$\textrm{M}$ direction, corresponding to the surrounding pseudogap regions in Fig.~\ref{fig:CP_model_final_5x5}, the dispersion curves have a smaller slope $v_\textrm{g}$ compared to the $\Gamma_0$--$\textrm{M}$ direction, probably due to an increased charge-paramagnon interaction strength, facilitated by the longer paramagnon coherence length. Furthermore, these modes are separated by a $0.05 \, \textrm{eV}$ `step', forming a `staircase', which might resemble the interaction of charge with one or multiple paramagnons.\\

\noindent
The disruption of the linear modes in Fig.~\ref{fig:antinodal_measurements_withLines} between $\Gamma_0$ and $\textrm{M}$ suggests that no charge dispersion across the plaquettes is possible beyond $0.25 \, \textrm{eV}$. The reason for this may be that (para)magnons become localized around this energy upon going from the underdoped (pseudogap) to the optimally doped regime, as shown by RIXS measurements such as Fig.~3 of \citeauthor{Braicovich2008} \cite{Braicovich2008} and Fig.~5e of \citeauthor{Guarise2014} \cite{Guarise2014}. Similar findings in \LSCO\ (LSCO) are found as well \cite{Dean2013_NatMat,LeTacon2011,LeTacon2011,Chaix2018}. The exact position of the broadened paramagnon peak at $\Gamma_0$ for Bi-2212 in the work from \citeauthor{Guarise2014} is not resolvable, but contributions at $0.25 \, \textrm{eV}$ for small $q_\parallel$ are possible, like in LSCO \cite{Dean2014}. As a consequence of this localized paramagnon, charge transport is eliminated above this energy in optimally doped regions.\\

\noindent
In the pseudogap phase ($\Gamma_1$--$\textrm{M}$), the paramagnon has a delocalised character \cite{Guarise2014,Braicovich2008}, therefore facilitates collective charge transport even above $0.25 \, \textrm{eV}$, which is evident in Fig.~\ref{fig:antinodal_measurements_withLines}. This behavior may be linked to the incoherent charge transport regime between $0.2$ and $1 \, \textrm{eV}$ which has been observed in optical conductivity measurements \cite{Marel2003,Heumen2022}.\\

\subsubsection{Charge transport}

\noindent
The dc and optical charge conductivity behavior as a function of temperature is essential for understanding the strange metal. \citeauthor{Heumen2009} \cite{Heumen2009} reported a correlation between the free-carrier optical conductivity and the `glue function' in the normal state, revealing a robust peak in the $50 - 60 \, \textrm{meV}$ range for all measured charge-carrier concentrations and temperatures up to $290 \, \textrm{K}$. The constructed glue function from the infrared optical conductivity represent an attractive interaction mediated by virtual bosonic excitations (lattice vibrations, spin-polarized or charge density fluctuations) in the solid. However, there is a significant difference between the underdoped and optimally doped samples since an enhancement in the electron-boson coupling in the energy range $0.2 - 0.3 \, \textrm{eV}$ occurs in the optimally doped sample.\\

\noindent
Inspired by Heumen and Zaanen \cite{Heumen2022}, we interpret the charge transport as consisting of two channels, the one near $\Gamma_0$ (Drude-like) and the other near $\Gamma_1$ (incoherent), which are parallel and together determine the charge transport. In our view, the Drude-like channel is eliminated above $0.25 \, \textrm{eV}$, while the incoherent channel is present at any energy. Therefore, the interpretation of the enhancement of the charge-boson coupling \cite{Heumen2009} can instead be regarded as a direct consequence of the elimination of the Drude-like channel. Namely, the Drude-like channel dominates the total charge transport below $0.25 \, \textrm{eV}$, whereas the incoherent channel takes over above this energy, giving rise to an apparent enhancement of the charge-boson coupling. This enhancement would then purely be a consequence of the coupling being stronger in the incoherent channel than in the Drude-like channel. This idea could also explain why such an enhancement is not observed in an underdoped sample in the pseudogap phase \cite{Heumen2009}. Namely, the Drude-like charge transport channel is absent (or less dominant) in such a sample, the electron-boson coupling does not show any drastic change around $0.2 - 0.3 \, \textrm{eV}$.\\

\noindent
The existence of two charge transport channels may also help to explain why the expected mirror symmetry around the Brillouin zone boundary in the $\textrm{M}$ point is not present in our measurements shown in Fig.~\ref{fig:antinodal_measurements}. Presumably, the incoming electrons scattering with charge in the coherent Drude-like channel have a smaller chance to undergo an Umklapp scattering process and thus the modes associated with the Drude-like channel will be strongly suppressed at higher-order $\Gamma$ points. On the contrary, this Umklapp scattering can take place more easily with electrons which scatter with the incoherent charge transport channel, which means that this mode exists not only at $\Gamma_0$, but also at $\Gamma_1$ and other higher-order $\Gamma$ points.\\

\subsubsection{Fluctuating stripes}

\noindent
As mentioned before, the linear dispersion of the propagating modes in the anti-nodal direction is smaller than calculated by \citeauthor{Eede2023} and \citeauthor{Guarise2014}. The complexity of the charge and (para)magnon or spin hybridization has been discussed by 
\citeauthor{Cvetkovic2007} from a different perspective, namely a quantum melting of crystals where topological defects (dislocations) have proliferated into quantum electronic liquid crystals of smectic kind \cite{Cvetkovic2006,Cvetkovic2007}. The resulting fluctuating stripes are characterized by an electronic shear velocity of the dislocations, which is of the order of the group velocity of the plasmon which has been calculated before \cite{Eede2023}.\\

\noindent
The fluctuating stripe interpretation rests on the characteristic disorder scales of the spin (paramagnon) fluctuation system, which should be rooted in the charge fluctuations. For an optimally doped cuprate, the characteristic length scale of the system is determined by the screening length of the delocalised dislocations, called $\lambda_\textrm{s}$, which is given by $\lambda_\textrm{s} = c_T / \Omega$. Here, $c_T$ is the transversal phonon velocity ($1 \, \textrm{eV} \, \si{\angstrom} / \hbar$) and $\Omega$ is the shear Higgs mass ($50 \, \textrm{meV}$) \cite{Cvetkovic2007,Tranquada2004}. From these estimate values, it follows that $\lambda_\textrm{s} = 2\, \textrm{nm}$, so in order to observe this massive shear photon mode an excellent energy resolution ($< 50 \, \textrm{meV}$) would be required, in combination with nanometer spatial resolution. This theoretically predicted mode by Cvetkovic and Zaanen has a group velocity $v_\textrm{g}$ of approximately $1.1 \times 10^{5} \, \textrm{m} \, \textrm{s}^{-1}$, which is on the order of the group velocity measured in our STEM-EELS measurements and furthermore our spatial resolution is approximately $1.6 \,\textrm{nm}$ as given by the probe size. Based on the spatial electronic liquid phase observed via STM by \citeauthor{Wise2008} in an optimally doped cuprate \cite{Wise2008}, the smectic electronic liquid forms a periodic array in one spatial direction \cite{Emery2000} which may represent a massive shear photon mode in Fig.~\ref{fig:antinodal_measurements_withLines} which Zaanen predicted as a signature of the existence of dynamical stripes in cuprate superconductors \cite{Cvetkovic2006}.\\

\noindent
In the liquid phase, at distances small compared to $\lambda_\textrm{s}$, the plasmon carries shear components at finite momenta just like a longitudinal phonon. These shear components should be separated from the plasmon, turning it into the purely compressional mode of the liquid at distances large compared to $\lambda_\textrm{s}$. This will cause a linear mode coupling between the plasmon and the massive shear photon, with the latter using some electromagnetic weight from the former.\\

\subsection{Concluding remarks}

\noindent
We have performed STEM-EELS measurements on the optimally doped cuprate \BSCCOtwo. Based on our proposed model, the $2 \times 2 \, \textrm{nm}^2$ area at which we measured probably contains part of an optimally doped plaquette and part of the pseudogap region. We observe traces of the strongly damped $1 \, \textrm{eV}$ plasmon mode, but in the lower energy region we find an additional set of linearly dispersing modes with different slopes in the anti-nodal direction. The three or so modes which appear near $\Gamma_0$ are ascribed to the massive shear photon predicted by Cvetkovic and Zaanen \cite{Cvetkovic2007} since the $v_\textrm{g}$ is similar to the theoretical prediction. These modes are cut off above $0.25 \, \textrm{eV}$ in our measurements, which illustrates the inflexibility of localized paramagnons in the optimally doped plaquettes, thereby reducing the charge degrees of freedom. Other dispersive modes start from $\Gamma_1$ and form a kind of `staircase' with steps of $0.05 \, \textrm{eV}$ which survives above $0.25 \, \textrm{eV}$. We suspect that these modes originate from the pseudogap regions where the delocalized paramagnon dispersion ensures that the charge fluctuations can persist at energies above $0.25 \, \textrm{eV}$. We conjecture that the $0.05 \, \textrm{eV}$ `steps' which are observed near $\Gamma_1$ represent the coupling between charge and one or multiple paramagnons.\\

\noindent
The charge transport is interpreted to consist of two channels, the one near $\Gamma_0$ (Drude-like) and the other near $\Gamma_1$ (incoherent), which are parallel and together determine the charge transport, in view of Heumen and Zaanen \cite{Heumen2022}. In our view, the Drude-like channel is eliminated above $0.25 \, \textrm{eV}$, while the incoherent channel is always present. Therefore the enhancement of the charge-boson coupling in the energy range $0.2 - 0.3 \, \textrm{eV}$ \cite{Heumen2009} can instead be regarded as a consequence of the elimination of the Drude-like channel and be merely a manifestation of the incoherent charge channel taking over above this energy. As such, it makes sense that this enhancement was not seen in an underdoped sample, since the Drude-like channel is absent.\\

\noindent
Our prediction that the Drude-like channel is absent in the pseudogap phase should be tested with future STEM-EELS experiments with nanometer and sub-$50 \, \textrm{meV}$ resolution on underdoped samples in the pseudogap phase.\\

\appendix
\section{Supplementary Material}

\subsection{EELS data acquisition} \label{supplementary:EELS_data_acquisition}

\noindent
Momentum resolved EELS experiments were performed on the monochromated, spherical aberration-corrected Nion UltraSTEM-100MC-Hermes dedicated STEM instrument, equipped with a Nion IRIS high resolution EELS spectrometer and a Dectris ELA direct electron detector for EELS. The microscope was operated at $60 \, \textrm{kV}$ and the energy resolution in the experiments was $15 \, \textrm{meV}$, as set by the monochromator slit. The momentum resolved EELS experiments were performed using a rectangular collection aperture \cite{Fossard2017,Qi2021}. Each EELS measurement was a multi-frame acquisitions of the spectrometer camera, subsequently aligned and averaged.\\

\noindent
The electron optics before and after the sample are adjusted to set the convergence semi-angle $\alpha$ and the collection semi-angle $\beta$, respectively. These angles determine the momentum resolution $\Delta q$ and spatial resolution $\Delta x$ (probe size) according to \cite{Hage2013,Egerton2007}
\begin{subequations} \label{eq:dq_and_dx}
\begin{align}
    \Delta q &\approx \frac{4 \pi}{\lambda}  \sin\frac{1}{2} \sqrt{\alpha^2 + \beta^2},\\
    \Delta x &\approx \frac{\lambda}{2\alpha},
\end{align}
\end{subequations}
where $\lambda \approx 4.9 \, \textrm{pm}$ is the wavelength of a $60$-$\textrm{keV}$ electron. Eq.~\ref{eq:dq_and_dx} is used to estimate $\Delta q$ and $\Delta x$ in the main text. The measurements in Fig.~\ref{fig:initial_measurements} were performed using a weakly convergent beam with $\alpha=\beta = 2 \, \textrm{mrad}$ (Fig.~\ref{subfig:STEM_2mrad_on_withDiffr}), and with a highly convergent beam with $\alpha = 30 \, \textrm{mrad}$ and $\beta = 44 \, \textrm{mrad}$ (Fig.~\ref{subfig:STEM_30mrad_on_withDiffr}). All measurements after that, in Figs.~\ref{fig:antinodal_measurements}--\ref{fig:nodal_measurements}, were done using $\alpha = 1.5 \, \textrm{mrad}$ and $\beta = 1 \, \textrm{mrad}$. The corresponding $\Delta q$ and $\Delta x$ are mentioned in the main text.\\

\newgeometry{
  top=4cm,          
  inner=1.5cm,
  outer=1.5cm,
  bottom=3cm,
  headheight=3ex,
  headsep=1cm,      
}

\addcontentsline{toc}{chapter}{References}
\printbibliography

\end{document}